\documentclass[12pt,twocolumn,english]{revtex4}
\usepackage{mathptmx}

\usepackage[T1]{fontenc}
\usepackage[latin9]{inputenc}
\usepackage{amssymb}
\usepackage{graphicx}
\usepackage{esint}

\makeatletter

\@ifundefined{textcolor}{}
{%
 \definecolor{BLACK}{gray}{0}
 \definecolor{WHITE}{gray}{1}
 \definecolor{RED}{rgb}{1,0,0}
 \definecolor{GREEN}{rgb}{0,1,0}
 \definecolor{BLUE}{rgb}{0,0,1}
 \definecolor{CYAN}{cmyk}{1,0,0,0}
 \definecolor{MAGENTA}{cmyk}{0,1,0,0}
 \definecolor{YELLOW}{cmyk}{0,0,1,0}
}

\usepackage{times}

\def\aap{{Astronomy and Astrophys.}}

\def\jgr{{J.~Geophys.~Res.}}
\def\mnras{{MNRAS}}

\def\apj{{\it Astrophys. J.\ }}
		
\def\apjl{{\apj\ \it Lett.\ }}

\def\pre{{\it Phys.~Rev.~E\ }}

\def\jgr{{\it J.~Geophys.~Res.\ }}

\def\physrep{{\it Phys.~Rep.\ }}
\def\mnras{{\it Mon. Not. R. Astron. Soc.\ }}

\usepackage{hyperref}

\makeatother

\usepackage{babel}
\begin{document}

\title{Proton-Helium Spectral Anomaly as a Signature of Cosmic Ray Accelerator}

\author{M.A. Malkov$^{1}$, P.H. Diamond$^{1}$, R.Z. Sagdeev$^{2}$ }

\affiliation{$^{1}$CASS and Department of Physics, University of California,
San Diego, La Jolla, CA 92093\\$^{2}$University of Maryland, College
Park, MD 20742-3280}
\begin{abstract}
The much-anticipated proof of cosmic ray (CR) acceleration in supernova
remnants (SNR) must hinge on full consistency of acceleration theory
with the observations; direct proof is impossible because of the orbit
scrambling of CR particles. The recent ATIC, CREAM and PAMELA experiments
indicated deviations between helium and proton CR spectra deemed inconsistent
with the theory, since the latter does not differentiate between elements
of ultrarelativistic rigidity. By considering an initial (injection-)
phase of the diffusive shock acceleration (DSA), where elemental similarity
does not apply, we demonstrate that the spectral difference is, in
fact, a unique signature\emph{ }of the DSA. Collisionless plasma SNR
shocks inject more He$^{2+}$relative to protons when they are stronger
and so produce harder helium spectra. The injection bias is due to
Alfven waves driven by the more abundant protons, so the He$^{2+}$
ions are harder to trap by these waves because of the larger gyroradii.
By fitting the p/He ratio to the PAMELA data, we bolster the DSA-case
for resolving the century-old mystery of CR origin.
\end{abstract}
\maketitle
Cosmic rays (CR), discovered in 1912 \cite{Hess12}, are subatomic
charged particles with a power-law energy spectrum extended up to
$\sim10^{20}$eV. At least to $\sim10^{15}$eV, they are commonly
believed to be accelerated by diffusive shock acceleration (DSA, or
Fermi-I \cite{Fermi49,axf77,krym77,Bell78,BlandOst78}) mechanism,
operating in supernova remnant (SNR) shocks (see \cite{BlandEich87,MDru01}
for a review). Recent precise measurements of proton and He$^{2+}$
spectra by PAMELA spacecraft \cite{Adriani11} indicate a small but
significant difference between the two, confirming earlier results
of ATIC \cite{ATIC2_09} and CREAM \cite{CREAM10,CREAM11}. Since
the DSA is electromagnetic in nature and accelerates all ultrarelativistic
species with equal rigidities alike, it was claimed inconsistent with
this difference. 

Indeed, at the basic level the DSA mechanism predicts a power-law
momentum distribution $\propto p^{-q}$ for the accelerated CR, where
the index $q$ depends on the shock Mach number $q=4/\left(1-M^{-2}\right)$.
Therefore, $q\approx4-4.1$ seems to be rigorous for strong shocks
($M\gg1$). At the same time, the subsequent escape from the Galaxy,
partial escape of CR from the shock in the course of acceleration,
and back-reaction of accelerated particles on the shock structure,
introduce deviations of observed spectra from the above power law.
Uncertainties in these corrections, not so much in the measurements,
prevent validation of the DSA as the mechanism for the CR production
in the Galaxy. 

Nevertheless, there is one fundamental property of this mechanism
that can be tested independently from the above uncertainties. It
is seen from the equations of particle motion in electric and magnetic
fields $\mathbf{E}$ and $\mathbf{B}$, written for the rigidity of
CR nucleus $\vec{\mathcal{R}}=\mathbf{p}c/eZ$, where $\mathbf{p}$
is the momentum and $Z$ is the charge number:

\begin{equation}
\frac{1}{c}\frac{d\vec{\mathcal{R}}}{dt}=\mathbf{E}\left(\mathbf{r},t\right)+\frac{\vec{\mathcal{R}}\times\mathbf{B}\left(\mathbf{r},t\right)}{\sqrt{\mathcal{R}_{0}^{2}+\mathcal{R}^{2}}},\label{eq:RigMotion}
\end{equation}

\begin{equation}
\frac{1}{c}\frac{d\mathbf{r}}{dt}=\frac{\vec{\mathcal{R}}}{\sqrt{\mathcal{R}_{0}^{2}+\mathcal{R}^{2}}}.\label{eq:coordMotioin}
\end{equation}
Here $\mathcal{R}_{0}=Am_{p}c^{2}/Ze$, with $A$ being the atomic
number. These equations show that if protons and He$^{2+}$ ions enter
the acceleration at $\mathcal{R}\gg\mathcal{R}_{0}$ in a certain
proportion $N_{p}/N_{{\rm He}}$, this ratio is maintained in course
of acceleration and the rigidity spectra are identical. Moreover,
if the both species leave (escape) the accelerator and propagate to
the observer largely without collisions, they will maintain the same
p/He ratio even if their individual spectra change considerably.

\begin{figure}
\includegraphics[bb=0bp 0bp 620bp 600bp,scale=0.7,angle=-90]{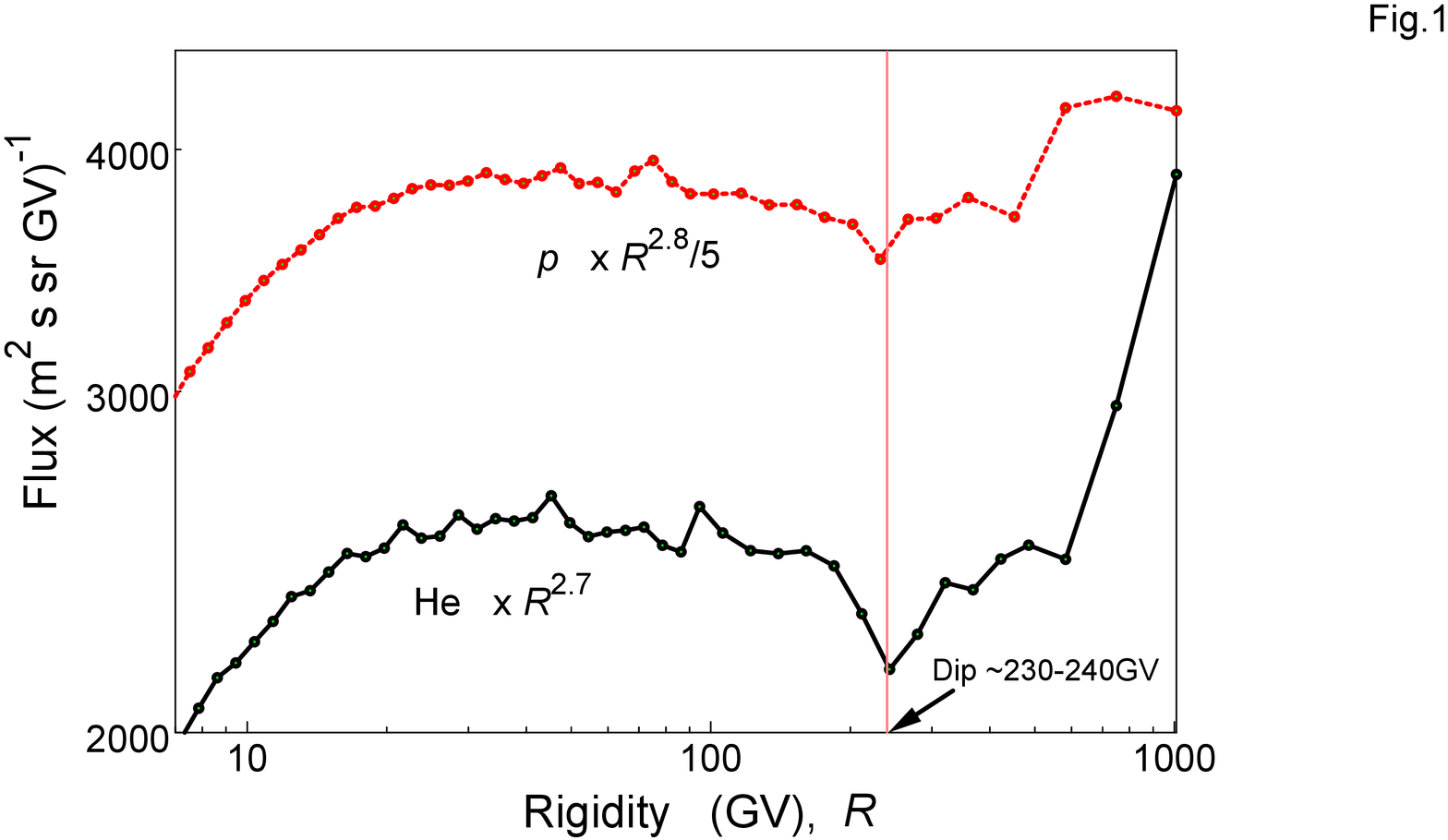}\caption{PAMELA fluxes of He and protons. He (solid line) and protons (dashed
line) are multiplied by $\mathcal{R}^{2.7}$ and $\mathcal{R}^{2.8}/5$,
respectively (proton spectrum artificially reduced to emphasize its
similarity with He spectrum). Circles represent PAMELA points adopted
from Supporting Online Material for \cite{Adriani11}. The sharp rise
of the He beyond $\mathcal{R}\simeq800$ GV is likely to be associated
with growing errors, since it does not match with the ATIC-2 and CREAM
\cite{ATIC2_08,CREAM10} data at $\mathcal{R}\gtrsim10^{3}$ GV. The
``zig-zags'' on each spectrum (also present at lower energies) are
well within the error-bars (not shown here).\label{fig:Pamela-fluxes}}
\end{figure}

The observations, however, were indicating for some time \cite{JACEE98,AMS_prot_00,AMS_He00,ATIC2_08,CREAM10}
that the spectrum of He may be somewhat harder (by $\Delta q=q_{p}-q_{{\rm He}}\lesssim0.1$)
than that of the protons over a wide range of rigidities $\mathcal{R}\gg\mathcal{R}_{0}$.
Recently, the PAMELA team \cite{Adriani11} determined $\Delta q=q_{p}-q_{{\rm He}}$
of $N_{p}/N_{{\rm He}}$ ratio as a function of rigidity with an unprecedented
accuracy, $\Delta q=0.101\pm0.001$ for $\mathcal{R}\gtrsim5$ GV,
where the finite $\mathcal{R}_{0}$ effect fades out %
\footnote{Possible secular effects from the small $\mathcal{R}_{0}^{2}/\mathcal{R}^{2}\ll1$
corrections in eqs.(\ref{eq:RigMotion}-\ref{eq:coordMotioin}) can
hardly be important due to the statistical nature of the observed
spectra. The \emph{stochastic instability} of the same element's orbits
is almost certainly more important for large $\mathcal{R}.$ A nearly
perfect isotropy of the observed CRs is a strong evidence for that.%
}. This finding challenges the DSA as a viable mechanism for galactic
CR acceleration. The challenge is best seen from a remarkable similarity
of the helium and proton spectra shown in an ``enhanced'' format,
in which \emph{p}-flux is multiplied by $\mathcal{R}^{2.8}$ and He-flux
by $\mathcal{R}^{2.7}$, Fig.\ref{fig:Pamela-fluxes}. 

While both spectra deviate from their power-laws, they do it synchronously
($N_{p}/N_{{\rm He}}$ is measured with significantly higher precision
than $N_{p}$ or $N_{{\rm He}}$, see below). First, let us focus
on the following three common features of the He and proton spectra:
(i) almost identical (three digits in the indices) convex shapes at
$5<\mathcal{R}<230-240$ GV with a likely roll-over towards the right
end of this interval (ii) sharp dip at $\mathcal{R}=230-240$ GV (iii)
upturn with nearly the same slope at $\mathcal{R}>230-240$ GV.

These features are clues for possible acceleration/propagation scenarios.
In particular, the He and proton spectra cannot come from independent
sources in their entireties. Otherwise, one is faced with the dip
coincidence and the overall shape similarity. Neither can they come
from a single shock, since the DSA and the subsequent propagation
are inconsistent with the spectral variations shown in the features
(i-iii). The remaining possibility seems to be that the low-energy
part ($\mathcal{R}<230-240$ GV) originates from one source (S1) while
the rest comes from the source(s) S2, including the invisible (under
the S1) part with $\mathcal{R}<230-240$ GV. S1 is likely to be a
local source with soft spectrum and a very low cut-off or a spectral
break. The source(s) S2 generates a harder, featureless spectrum that
merges into (or comprises) the galactic background (see, however,
\cite{VladimirMoskPamela11} for more scenarios). 

Despite considerable differences between the putative sources S1 and
S2, the $p$/He ratio is a remarkably featureless function of rigidity,
$\propto\mathcal{R}^{-0.1}$, in a wide rigidity range including the
transition zone, at $\mathcal{R}=230-240$ GV (see Ref.\cite{Adriani11}
and below). This points at a common (for S1 and S2 and intrinsic to
the DSA) mechanism that should account for the same 0.1 difference
in independent sources. By virtue of eqs.(\ref{eq:RigMotion}-\ref{eq:coordMotioin}),
such a difference cannot arise in the region $\mathcal{R}\gg\mathcal{R}_{0}$.
Therefore, it must originate at $\mathcal{R}\ll\mathcal{R}_{0}$,
as we believe, in the following way. 

A small fraction of thermal upstream particles, after crossing the
shock may become subject to the DSA (to be ``injected'') if they
recross the shock in the upstream direction \cite{mv95}. Their amount
depends on shock obliquity and Mach number (we will focus on quasiparallel
shocks as more favorable for injection and further acceleration, but
the results can be extended to the field inclinations w.r.t. the shock
normal $\vartheta_{nB}\sim30-40^{\circ}$ \cite{SpitkovskyHybr11}
). 

\emph{In situ} observations \cite{Lee82} of the Earth's bow-shock
indicate that about $10^{-3}$ of incident protons are injected. It
is also known from such observations that, on average, 1.6 more ${\rm He^{+2}}$
ions than protons are injected \cite{IpavichHe_H84}. This He/p injection
excess does not explain the PAMELA He/\emph{p} excess unless it grows
with the shock Mach number when the latter increases to the SNR-scales
($M\sim100$). This is not known from \emph{in situ} observations
of shocks limited to Alfven Mach numbers $M_{A}\sim M\sim5$. Therefore,
we use the injection model \cite{m98} that predicts such growth.
It is consistent with the observations \cite{IpavichHe_H84} at low
Mach numbers and with the recent simulations \cite{SpitkovskyHybr11}
in the important for the He hardening range of $M_{A}\sim5-30$. 

The mechanism of preferential He injection is based on the larger
He gyroradius downstream. Upon crossing the shock, both protons and
He randomize their downstream frame velocity, which is $\simeq V_{{\rm s}}\left(1-1/r\right)$
(where $V_{{\rm s}}$ is the shock velocity and $r$ is its compression
ratio) by interacting with magnetohydrodynamic waves, predominantly
driven by the protons. We may consider the waves to be frozen into
the flow since $M_{A}=V_{{\rm s}}/C_{A}\gg1$, where $C_{A}$ is the
Alfven speed. As the proton gyroradius is a half of that of He$^{2+}$
(for the same velocity $\sim V_{{\rm s}}$), the Helium ions have
better chances to return upstream since protons are retained by the
downstream waves more efficiently. According to the model, the injection
rates of both species decrease with $M_{A}$ but the proton injection
decreases faster. 

To quantify this effect, the model admits an initially unknown fraction
of incident protons to return upstream where they drive a nearly monochromatic
Alfven (magneto-sonic) wave. After being amplified by shock compression
and convected further downstream, the wave traps most of the protons
and regulates their return upstream. (He$^{+2}$ ions are still regarded
as test-particle minority). The monochromaticity of the wave upstream
is justified by the narrowness of the escaping beam distribution compared
to its bulk velocity upstream. The wave amplitude settles at a predictable
level due to the obvious self-regulation of proton escape: if the
escape is too strong, the wave grows to trap more protons.

\begin{figure}
\includegraphics[scale=0.7,angle=-90]{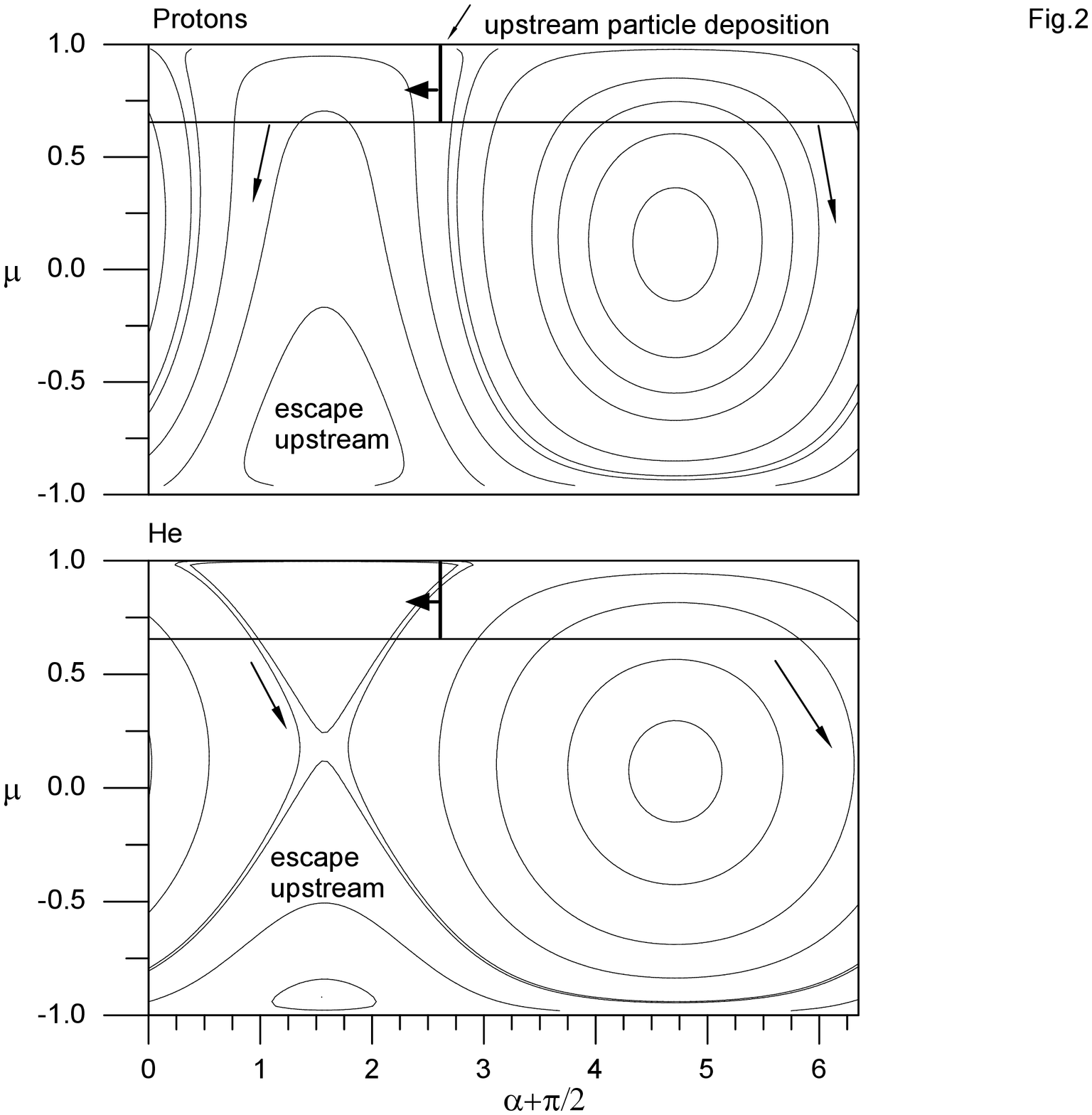}\caption{Protons (top panel) and He (bottom panel) in a monochromatic wave
downstream. The trajectories ($H={\rm const}$, eq.{[}\ref{eq:Ham}{]})
are shown on the particle phase plane $\alpha,\mu$ for the same particle
velocity $V$, yielding $v=1.2$ for protons and $v=2.4$, for He.
The wave amplitude $B_{\perp}/B_{0}=4$, which corresponds to the
wave amplitude $B_{\perp}\simeq B_{0}$ saturated upstream and compressed
later by the shock. The vertical bars at the top of each panel schematically
show the particle entrance from the shock surface when it moves to
the left across the phase plane.\label{fig:Phase-space-of}}
\end{figure}

The mechanism is illustrated by Fig.\ref{fig:Phase-space-of}, where
particle trajectories in the downstream wave are depicted in coordinates
$\mu=V_{\parallel}/V$ (cosine of the pitch angle w.r.t. the average
magnetic field $\mathbf{B}_{0}$) and $\alpha=k_{2}z+\phi$, where
$k_{2}$ is the wave number downstream (related to that of the upstream
by $k_{2}\approx rk_{1}$), $z$ is the coordinate (directed downstream)
parallel to $\mathbf{B}_{0}$ and shock normal, and $\phi$ is the
gyrophase. Particles enter the downstream phase plane at its top when
the shock sweeps in the negative $\alpha$ direction. Then they begin
to move in the downstream wave along the lines of constant Hamiltonian

\begin{equation}
H=\sqrt{1-\mu^{2}}\cos\alpha+\frac{1}{2}v\mu^{2}-\frac{B_{0}}{B_{\perp}}\mu\label{eq:Ham}
\end{equation}
where $B_{\perp}$ is the wave amplitude and $v=k_{2}V/(eZB/Am_{p}c)$.
For the same particle velocity $V$ (which is an integral of motion),
the parameter $v$ for He, $v_{_{{\rm He}}}=2v_{p}$, which makes
the escape zone on the phase plane larger and more accessible to He$^{2+}$
ions than to protons. Note that in order to escape upstream, particles
should cross the lines $H=const$ which is enabled by perturbations
\footnote{In computing the escape rate from the downstream side, the model specifically
assumes that particles are evenly distributed on each isoenergetic
($V={\rm const}$) surface (ergodicity assumption). This is a good
approximation for particles escaping from far downstream, where the
shock thermalization is completed and particle distribution depends
only on integrals of motion (energy). For particles escaping soon
after crossing the shock, in which case the particle-shock interaction
can be regarded as reflection rather than leakage, the ergodic \emph{Ansatz}
may become less accurate and the reflection process requires further
study. The reflection/leakage dichotomy is often emphasized in simulation
analyses but it has not been substantiated by establishing specific
criteria. A simple such criterion is to regard particle return from
the first wave period downstream as reflection while that from the
second or more distant periods as leakage. %
}

For $M_{A}\gg1$, the injection is suppressed according to $\eta_{p}\propto M_{A}^{-1}\ln\left(M_{A}/M_{*}\right)$,
where $M_{*}\sim10$. The injection is more efficient at smaller $M_{A}$,
but its dependence upon $M_{A}$ is complicated \cite{m98}. A formal
fit to the proton injection suppression factor gives $\eta_{p}\approx0.4\cdot M_{A}^{-\sigma_{p}}$
with $\sigma_{p}\approx0.6$. The He$^{2+}$ injection is suppressed
to lesser extent, yielding $\eta_{{\rm He}}\approx0.5\cdot M_{A}^{-\sigma_{{\rm He}}}$
with $\sigma_{{\rm He}}\approx0.3$. Both scalings are valid in the
range $5\lesssim M_{A}\lesssim100$ so that the assumption about the
test particle dynamics of He$^{2+}$ applies even for $M_{A}\gtrsim100$,
where $\sigma_{p}$ must be larger, according to the $\eta_{p}\propto M_{A}^{-1}\ln\left(M_{A}/M_{*}\right)$
asymptotic result. To accommodate this trend, we adjust $\sigma_{p}$
within the range $0.6-0.9$ and $\sigma_{{\rm He}}$ within $0.15-0.3$.
Note, however, that very high $M_{A}$, where the index $\sigma_{p}$
grows, are linked with small SNR radii and their contribution is less
important.

From an SNR lifetime, we therefore select the Sedov-Taylor phase as
the most important for the background CR production. The shock radius
grows with time as $R_{{\rm s}}\simeq C_{{\rm ST}}t^{2/3}$, where
$C_{{\rm ST}}=\left(2.03E/\rho_{0}\right)^{1/5}$, $E$ is the SN
energy and $\rho_{0}$ is the ambient density \cite{McKeeTruelove95}.
The shock speed is thus $V_{{\rm s}}=\left(2/5\right)C_{{\rm ST}}^{5/2}R_{{\rm s}}^{-3/2}$.
When the shock radius increases from $R_{{\rm min}}$ to $R_{{\rm max}}$,
the following number of CRs (with momentum $p$) are deposited in
the shock interior

\begin{equation}
N_{\alpha}\left(p\right)=A\intop_{M_{{\rm max}}^{-2}}^{M_{{\rm min}}^{-2}}f_{\alpha}\left(p,M\right)dM^{-2}\label{eq:CRdepos}
\end{equation}
where $M$ is the current shock Mach number, $M=V_{{\rm s}}/C_{{\rm s}}$,
$\alpha=p,{\rm He}$; $C_{{\rm s}}$ is the speed of sound and the
constant $A$ is not important since we are interested only in the
\emph{p}/He ratio. The spectra can be represented as follows

\begin{equation}
f_{\alpha}\propto\eta_{\alpha}\left(M\right)\left(\mathcal{R}_{{\rm inj}}/\mathcal{R}\right)^{q\left(M\right)}\label{eq:falfa}
\end{equation}
Here $\mathcal{R}_{{\rm inj}}$ is a reference (injection) rigidity,
which can be arbitrarily fixed at $\mathcal{R}_{{\rm inj}}=1$ GV,
since we are only concerned with the spectrum behavior at $\mathcal{R}\gg\mathcal{R}_{{\rm inj}},\mathcal{R}_{0}$. 

Introducing a new variable $x=4\ln$$\left(\mathcal{R}/\mathcal{R}_{{\rm inj}}\right)$,
using the integration variable $t=M^{-2}$ instead of $M$ and substituting
$q=4\left(1-M^{-2}\right)$, $\eta_{\alpha}\propto M^{-\sigma_{\alpha}}$,
for the \emph{p}/He ratio we obtain

\begin{equation}
N_{p}/N_{{\rm He}}=C\frac{\intop_{a}^{b}t^{\sigma_{p}/2}e^{-x/\left(1-t\right)}dt}{\intop_{a}^{b}t^{\sigma_{{\rm He}}/2}e^{-x/\left(1-t\right)}dt}\label{eq:PtoHe}
\end{equation}
where the constant $C$ is determined by the ratio of p\emph{/}He
concentrations. We also denoted $a=M_{{\rm max}}^{-2}\ll1$ and $b=M_{{\rm min}}^{-2}\lesssim$1. 

\begin{figure}
\includegraphics[bb=0bp 0bp 612bp 600bp,scale=0.7,angle=-90]{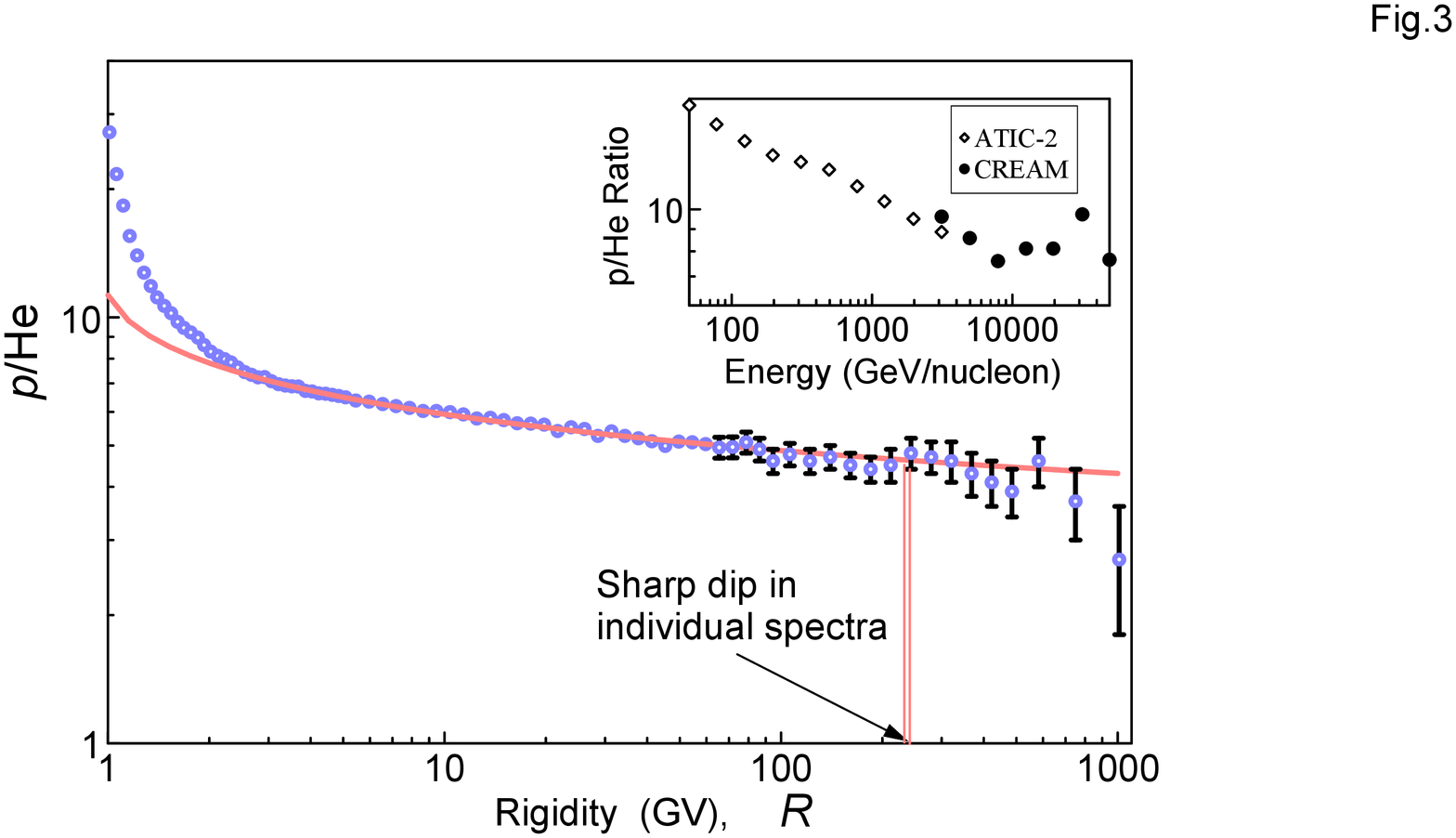}\caption{Main graph: PAMELA points (circles) and eq.(\ref{eq:PtoHe}) fit (line).
The fit is obtained for $\sigma_{p}=0.85,\;\sigma_{{\rm He}}=0.15$,
$M_{{\rm min}}=1.05$, $M_{{\rm max}}=100$ and the normalization
constant $C=15.5$ to match with the PAMELA data. This value is, however,
consistent with the 0.1 He abundance. The twenty highest rigidity
points are shown with the error-bars (stat.+syst.), where they seem
to become significant and the rightmost point clearly deviates from
the theoretical prediction (the data points adopted from the supporting
online material of \cite{Adriani11}). The proton and He spectral
breaks (see also Fig.\ref{fig:Pamela-fluxes}), collocated (within
uncertainties) at 230-240 GV, are shown with two vertical lines. At
higher rigidities the data from ATIC-2 \cite{ATIC2_09} and CREAM
\cite{CREAM11} are shown in the inset, however, as a function of
energy per nucleon (both adopted from \cite{CREAM11}). \label{fig:PtoHPamelaANDFIT}}
\end{figure}

The result given by eq.(\ref{eq:PtoHe}) is shown in Fig.\ref{fig:PtoHPamelaANDFIT}
along with the PAMELA \emph{p}/He ratio. The agreement is very good
besides the low rigidity range $\mathcal{R}\lesssim\mathcal{R}_{0}$
where it is not expected as the solar modulations, some further details
of injection \cite{m98}, and possible but largely unknown propagation
effects are not included in eq.(\ref{eq:PtoHe}). Therefore, we make
no attempts at fitting the $\mathcal{R}\lesssim\mathcal{R}_{0}\sim\mathcal{R}_{{\rm inj}}$
range in Fig.\ref{fig:PtoHPamelaANDFIT}, so the validity range of
the fit, $\mathcal{R}^{2}\gg\mathcal{R}_{0}^{2}$, i.e., $\mathcal{R}>2-3$
GV, is clearly seen from the plot. The deviation from the highest
rigidity point is likely to be due to large measurement errors and,
in part, due to the breakdown of $\eta_{\alpha}\propto M^{-\sigma_{\alpha}}$
scalings. 

On representing Eq.(\ref{eq:PtoHe}) as 

\[
N_{p}/N_{{\rm He}}=C\frac{F\left(\sigma_{p},x\right)}{F\left(\sigma_{{\rm He}},x\right)},
\]
for moderately large $x=4\ln\left(\mathcal{R}/\mathcal{R}_{{\rm inj}}\right)$,
we may obtain for $F$ 

\[
F\approx x^{\sigma_{\alpha}/2}e^{-x}\left\{ \Gamma\left(\nu\right)\left[1-\nu\left(\nu+1\right)\frac{1}{x}\right]-\frac{a^{\sigma_{\alpha}/2}}{\nu}\right\} ,
\]
where $\nu=\sigma_{\alpha}/2+1$ and $\Gamma$ denotes the gamma function.
The last term in the braces, that corresponds to the contribution
from highest Mach numbers, may be neglected, as $a\ll1$. For sufficiently
large $\mathcal{R}$, the \emph{p}/He ratio behaves as the following
power-law in $\ln\left(\mathcal{R}\right)$

\begin{equation}
N_{p}/N_{{\rm He}}\propto\left[\ln\left(\mathcal{R}/\mathcal{R}_{{\rm inj}}\right)\right]^{-\left(\sigma_{p}-\sigma_{{\rm He}}\right)/2}\label{eq:PtoHePL}
\end{equation}

The \emph{p}/He ratio\emph{ }at ultrarelativistic rigidities\emph{,}
as opposed to the\emph{ }individual spectra,\emph{ }is not affected
by the CR propagation, if collisions are negligible. Therefore, it
should be examined for telltale signs intrinsic to the particle acceleration
mechanism. The precise measurements of this ratio by the PAMELA \cite{Adriani11},
suggests reproducing their results theoretically with no free parameters.
While we have obtained a convenient control parameter for this quantity,
$\sigma\left(M_{A}\right)=\sigma_{p}-\sigma_{{\rm He}}$, from a collisionless
shock model best suited to the PAMELA rigidity range, the model predictions
need to be extended and improved systematically. Even though collisionless
shocks is a difficult subject of plasma physics, still not understood
completely \cite{Sagdeev79a,Kennel85,Papad85}, we expect modern simulations
\cite{Scholer02,SpitkovskyHybr11} to refine the proposed mechanism.
This will extend the theory's fit to a broader range spectrum, currently
being measured by the AMS-02, and help to determine whether or not
galactic CRs are produced in SNRs.

To conclude, there are alternative interpretations of the He/p spectral
hardening: (a) different SNR-type to contribute to the CR spectrum
\cite{Bier95,Zatsepin06,Adriani11}, (b) variable He/p concentration
in SNR environments \cite{DruryEscape11,Ohira11} and (c) CR spallation
\cite{BlasiChemComp11}. They are reviewed in \cite{VladimirMoskPamela11},
where it is pointed out that the overall data are best reproduced
if harder He spectra are directly released from accelerators.
\begin{acknowledgments}
We are indebted to the anonymous referees for helpful suggestions.
Support by the US DoE, Grant No. DE-FG02-04ER54738 is also gratefully
acknowledged.
\end{acknowledgments}
\clearpage

\clearpage

\clearpage

\end{document}